\documentclass[twocolumn,showpacs,preprintnumbers,amsmath,amssymb,aps,dvips,prb]{revtex4}
\usepackage{amsmath}
\usepackage[dvips]{graphicx}
\usepackage{dcolumn}
\newcolumntype{d}[1]{D{.}{.}{#1}}
\usepackage{bigdelim}
\usepackage{bm}
\usepackage{array}
\usepackage{longtable}
\usepackage{lscape}
\usepackage{float}
\usepackage{rotating}
\usepackage{multirow}
\usepackage{setspace}

\makeatletter

\newcommand{\Rmnum}[1]{\expandafter\@slowromancap\romannumeral #1@}
\makeatother

\begin{document}

\preprint{APS/123-QED}

\title{A novel nonlinear spin wave theory for the spin $\frac{1}{2}$ antiferromagnetic Heisenberg model on a triangular lattice}
\author{Lihua Wang$^{1}$}
\author{Sung Gong Chung$^{2,3}$}

\affiliation{$^1$
Computational Condensed-Matter Physics Laboratory, RIKEN, Wako, Saitama 351-0198, Japan}
\affiliation{$^2$
Department of Physics and Nanotechnology Research and Computation Center, Western Michigan University, Kalamazoo, Michigan 49008, USA  
}%

\affiliation{$^3$
Asia Pacific Center for Theoretical Physics, Pohang, Gyeonbuk 790-784, South Korea}

\date{\today}

\begin{abstract} 
We extend the nonlinear spin wave theory (NLSWT) for the spin 1/2 antiferromagnetic Heisenberg model on a triangular lattice (TAFHM). This novel NLSWT considers the corrections one order higher in $1/S$ than the linear spin wave theory (LSWT). It also distinguishes in which circumstance the negative energy excitation, the sign of the breakdown of LSWT, shall be renormalized to be positive both by a boson normal ordering and a self-consistent iteration. We draw a phase diagram by testing the stability of various magnetic orders for different parameters. In particular, the incommensurate configuration is found unstable by our study. The new phase transition point (PTP) of the collinear configuration agrees well with various previous studies. 
\end{abstract}
\maketitle

\section{\label{sec:LSW}Introduction}
$S=1/2$ TAFHM is extensively studied. This model may describe the low-energy physics of various quasi-2D triangular materials such as Cs$_2$CuCl$_4$~\cite{coldea}, $\kappa$-(BEDT-TTF)$_2$Cu$_2$(CN)$_3$~\cite{kanoda}, and EtMe$_3$Sb[Pd(dmit)$_2$]$_2$~\cite{kato} for which experiments have indicated a magnetically disordered Mott insulator, i.e., a possible spin liquid state, at low temperatures. Various theories are applied to understand how the geometric frustration and low dimensionality affect the physics. The diagonalization method \cite{bernu} is exact but limited in the lattice size, while the finite size effect is so strong to even change the phase diagram \cite{wang1}. There are also nearly-exact numerical simulations such as DMRG \cite{MQWeng}. They have encountered problems with overwhelmingly large entanglement when the dimensionality increases to 2D, although our own numerical work using the Entanglement Perturbation Theory(EPT), moves a small step further in this direction. The numerical simulation is powerful as we already see very rich details of the physics which is believed to be closely related to the geometric frustration, low dimensionality and strong quantum fluctuation. An example is the phase change due to the finite-size effect mentioned above.

However, currently, the most efficient theoretical framework to understand the model {\it in the thermodynamic limit} is the spin wave theory (SWT), which so far unfortunately only gives approximate description. There are two well known questions on the validity of SWT, i.e.,  if the assumed magnetic order is physical and if its approximation is sufficient when the first answer is positive, because the conversion to the bosonic Hamiltonian in an expansion of Holstein-Primakoff transformation \cite{primakoff} restores the spin statistics only either in large S limit or when higher order corrections are properly taken into account. To get an insight into the answers, one usually examines the excitation spectrum. First, there must be Goldstone modes in the K-space to reflect the symmetry of both the lattice and the assumed order. It is automatically satisfied in most SWT including LSWT because the unitary transformation, that is the starting point of SWT, naturally inherits those symmetries. Special cares, however, should be taken if it is involved with a self-consistent iteration or when there is a singularity cancelling issue in the perturbation. Second, the excitation spectrum must be positive. The emergence of the negative energy is a sign of the overwhelming quantum fluctuation over the assumed classical configuration, implying the breakdown of SWT. Furthermore the interplay of the above two factors can sometimes lead to subtle situations. Particularly for a spin $1/2$ TAFHM, LSWT has a series of continuous incommensurate spiral orders \cite{trumper} that yield a legal spectrum, but are unphysical at all. In contrast, the spectrum from LSWT has areas of negative energy for the collinear configuration when $1.2<\mu\equiv J'/J <2.0$, $\mu$ being the anisotropy defined below. But a variety of previous studies \cite{MQWeng, yunoki, bernu, wang1} suggest that the collinear order is realized in this regime. It is believed that the instability of LSWT manifests the insufficiency of approximation. Therefore one of the objectives of this paper is to see if a NLSWT can fill this gap, with the negative energy excitation being considered in a novel way. 

\begin{figure}
\begin{center}
\includegraphics[height=1.8in, width=2.in]{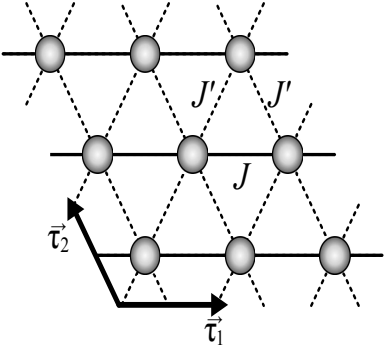}
\caption{\label{fig:lattice}Spin 1/2 TAFHM with the  anisotropic nearest neighbor couplings $J$ and $J'$ ($J'/J\ge 0$) indicated in the figure. The primitive translation vectors are denoted by $\tau_1$ and $\tau_2$. The periodic boundary conditions are assumed.}
\end{center}
\end{figure}

Now let us clarify the model. $J'$ mentioned above is the inter-chain spin exchange strength and $J$ intra-chain. See Fig.\ref{fig:lattice} for the illustration. The nearest neighbors along the three orientations, $0^o$, $60^o$ and $120^o$, are defined as $\left(i,j\right)_\delta$. $\delta_1$ and $\delta_3$ respectively correspond to the primitive translation vectors $\vec{\tau}_1$ and $\vec{\tau}_2$ shown in Fig.\ref{fig:lattice}. $\vec{x}$ direction is along $\vec{\tau}_1$.   
The Hamiltonian is 
\begin{equation}
\label{H1}
H=\sum_{\delta}{J_\delta \sum_{\left(i,j\right)_{\delta}}{\left(S_i^xS_j^x+S_i^yS_j^y+S_i^zS_j^z\right)}}
\end{equation}
Here we consider the anisotropy $J_1=J$, $J_2=J_3=J'$.

\begin{figure}
\begin{center}
\begin{tabular}{cc}
(a) $\mu=1.0$ & (b) $\mu=1.3$\\
\includegraphics[height=1.6in, width=1.6in]{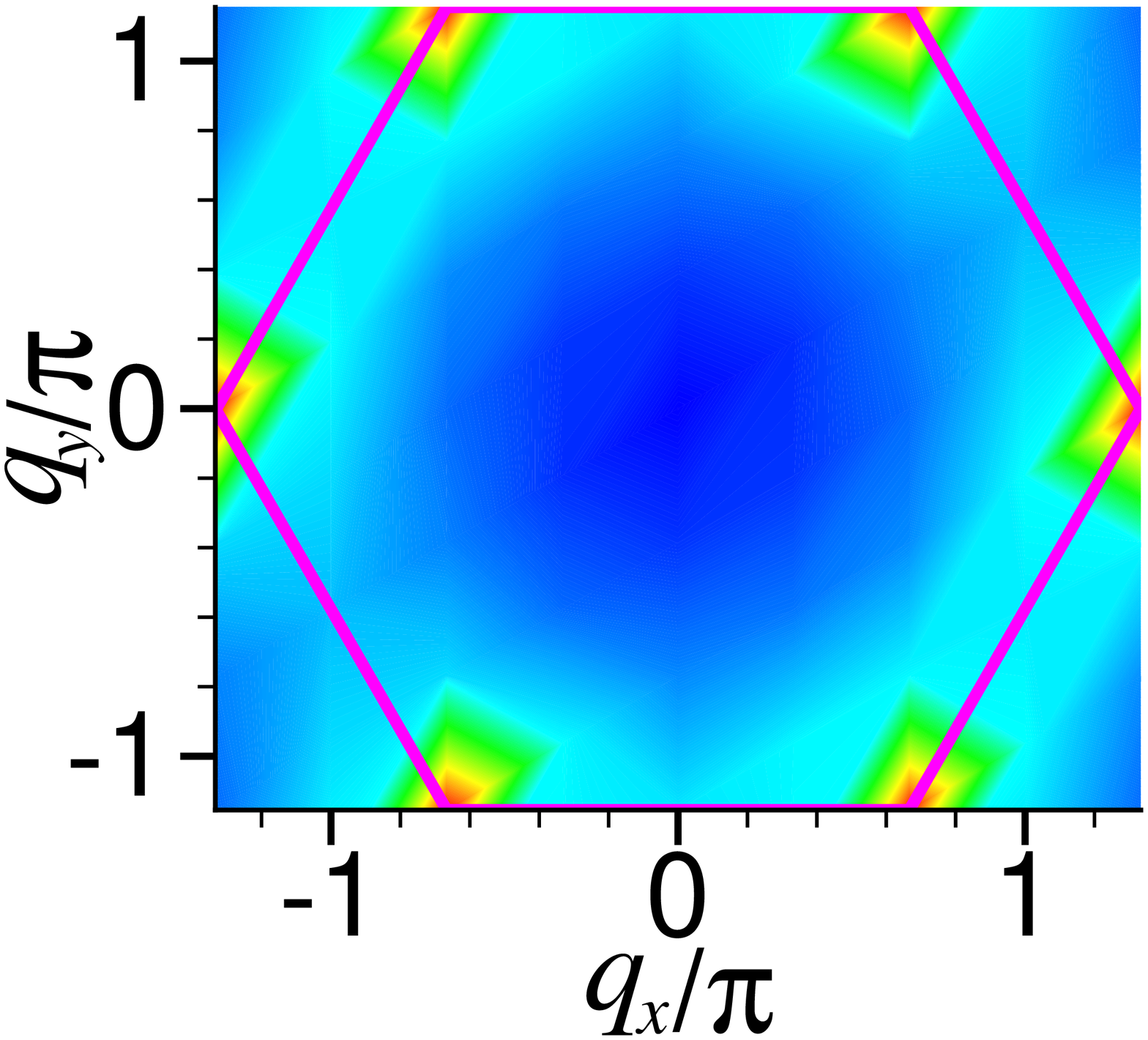} & \includegraphics[height=1.6in, width=1.6in]{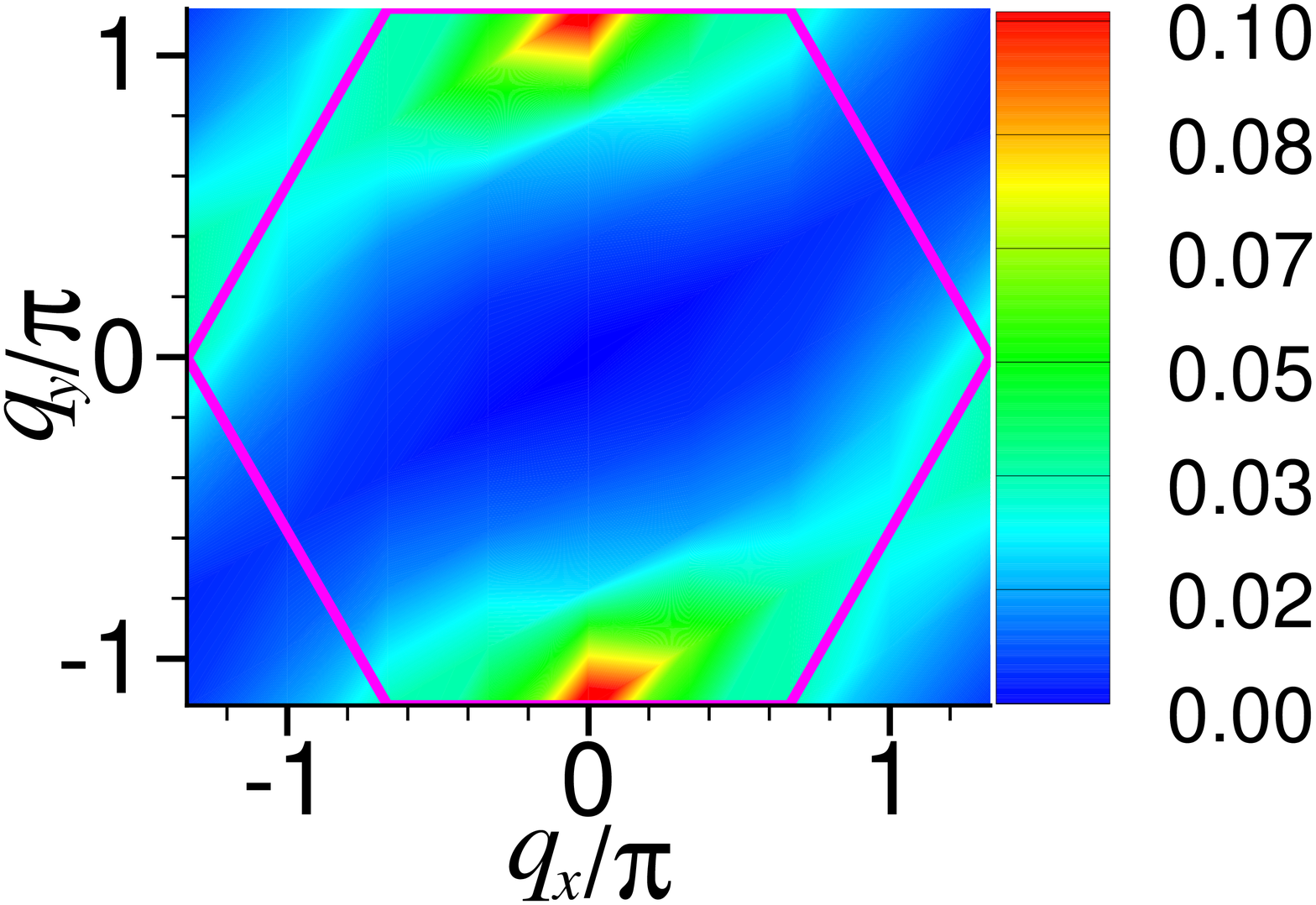}\\
 \end{tabular}
\caption{\label{Fig:spinstructure}Contour plots of the pin structure factor by EPT for a $6\times 6$ TAFHM (a) Isotropic when $\mu=1.0$ and (b) anisotropic $\mu=1.3$. (a) refers to the $120^o$ spiral spin configuration, (b) collinear.}
\end{center}
\end{figure}

The expected magnetic orders are signaled by the contour peaks at wave number $\vec{Q}$'s in the spin structure factor, Fig.\ref{Fig:spinstructure}, from our numerical simulation \cite{wang1}. Typically (a) $\vec{Q} =\left(\frac{4\pi}{3},0\right)$, etc, for the $120^o$ spiral configuration and (b) $\vec{Q} =\left(0,\frac{2\pi}{\sqrt{3}}\right)$, etc, collinear. They are used to rotate the laboratory coordinate frame to unify the direction of the local spin in SWT. 

Since many works \cite{capriotti,trumper,chernyshev,merino} have been done to formulate SWT for the spin 1/2 TAFHM, we simply start from the unitarily transformed Hamiltonian

\begin{align}
\label{H2}
H'=& \sum_{\delta}J_\delta \sum_{\left(i,j\right)_{\delta}}\bigl[S_i^yS_j^y+cos\left(\vec{Q}\cdot \vec{\delta}\right)\left(S_i^zS_j^z+S_i^xS_j^x\right)\notag\\
& +sin\left(\vec{Q}\cdot \vec{\delta}\right)\left(S_i^zS_j^x-S_i^xS_j^z\right)\bigr]
\end{align}

It is the Holstein-Primakoff transformation that transforms the above spin Hamiltonian to the bosonic operator (magnon) version. In practice it is approximated by expansion in $1/S$. For example, LSWT is in the order of $O(\left(1/S\right)^{-1/2})$. The expansion restores the spin statistics in large $S$ limit even in LSWT. However, for finite $S$, it needs infinite order to do so. In particular for $S=1/2$, LSWT is regarded as an insufficient expansion for certain combination of spin configuration and coupling parameter. Our NLSWT goes beyond LSWT one order higher in $1/S$. It takes the following expansion
\begin{align}
\label{PMT1}
S^x & = \sqrt{2S}\left(\frac{a^{\dagger}+a}{2}-\frac{a^{\dagger}a^{\dagger} a+a^{\dagger}aa}{8S}\right) \notag \\ 
S^y & = i\sqrt{2S}\left(\frac{a^{\dagger}-a}{2}-\frac{a^{\dagger}a^{\dagger} a-a^{\dagger}aa}{8S}\right)  \\
S^z & = S-a^{\dagger}a \notag
\end{align}
The Hamiltonian is rewritten in the K-space after the Fourier transformation as 
\begin{equation}
\label{H3}
H'=H_0+H_2+H_3+H_4
\end{equation}
where $H_0$ is a constant contribution to the ground state energy (GSE)

\begin{align}
\label{GSE0}
E_{GS}^0 \equiv H_0 = \frac{LJ}{4}cos{Q_x}+\frac{LJ'}{2}cos{\frac{Q_x}{2}}cos{\frac{\sqrt{3}Q_y}{2}}
\end{align}
with $L$ being the lattice size. On the other hand, $H_2$ is the quantum fluctuation bilinear with respect to magnon operators. It is

\begin{align}
\label{LH1}
H_2 = \sum_k{\left[A_k a_k^{\dagger}a_k-\frac{1}{2}B_k\left(a_k^{\dagger}a_{-k}^{\dagger}+a_{-k}a_k\right)\right]}
\end{align}
with

\begin{align}
\label{AKBK1}
A_k =& -Jcos{Q_x}-2J'cos{\frac{Q_x}{2}}cos{\frac{\sqrt{3}Q_y}{2}}+J\frac{cos{Q_x}+1}{2}cos{k_x}\notag\\
&+J'\left(cos{\frac{Q_x}{2}}cos{\frac{\sqrt{3}Q_y}{2}}+1\right)cos{\frac{k_x}{2}}cos{\frac{\sqrt{3}k_y}{2}} \\
B_k =& -J\frac{cos{Q_x}-1}{2}cos{k_x}\notag\\
&-J'\left(cos{\frac{Q_x}{2}}cos{\frac{\sqrt{3}Q_y}{2}}-1\right)cos{\frac{k_x}{2}}cos{\frac{\sqrt{3}k_y}{2}}
\end{align}

For the moment we focus on $H_0$ and $H_2$, which are the only terms considered by LSWT. $H_3$ and $H_4$ are left for the later discussion. There is a handy transformation named after Bobolyubov to diagonalize $H_2$. Its usual definition is as follows

\begin{align}
\label{BT1}
a_k &= u_k b_k +v_k b_{-k}^{\dagger} \notag\\
a_k^{\dagger} &= u_k b_k^{\dagger} +v_k b_{-k}
\end{align}
The coefficients, $u_k$ and $v_k$, are the roots of the following equations
\begin{align}
\label{BT2}
u_k^2 +v_k^2 &= \frac{S\left(A_k\right)A_k}{\sqrt{\left|A_K^2-B_K^2\right|}} \notag\\
2u_kv_k &= \frac{S\left(A_k\right)B_k}{\sqrt{\left|A_K^2-B_K^2\right|}} 
\end{align}
Note that $S\left(A_K\right)\equiv Sign\left(A_k\right)$ is used in (\ref{BT2}). Since (\ref{BT2}) is in second order, one usually complements it with the following convention
\begin{equation}
\label{parity1}
u_k^2-v_k^2=1
\end{equation} 
to preserve the same commutation relationship of magnon. Then $H_2$, after the Bogolyubov transformation, reads as
\begin{equation}
\label{LH2}
H_2=-\sum_{k}{\frac{A_k}{2}}+\sum_k{\epsilon_k\left(b_k^{\dagger}b_k+\frac{1}{2}\right)}
\end{equation}
where
\begin{equation}
\label{lepsilon1}
\epsilon_k=S\left(A_k\right)S\left(A_k^2-B_k^2\right)\sqrt{\left|A_k^2-B_k^2\right|}
\end{equation}
defines the excitation spectrum. Since the ground state is the vacuum of Bogolyubov bosons, $H_2$ contributes to GSE as follows
\begin{equation}
\label{GSE2}
E'_{GS}=\frac{1}{2}\sum_k{\left(\epsilon_k-A_k\right)}
\end{equation}
Therefore GSE calculated within LSWT is 
\begin{align}
\label{EQ:GSE1}
E_{GS}=E^0_{GS}+E'_{GS}
\end{align}

Now look at the property of $\epsilon_k$. It can be easily shown that $A_k=B_k$ at $\vec{k}=\left(0,0\right)$, and $A_k=-B_k$ at $\vec{k}=\pm \vec{Q}$. Therefore they are always the zero-energy points, the so-called Goldstone modes. It is the characteristics of a system which spontaneously breaks the symmetry in the thermodynamic limit. In addition, the fisrt derivative of a linear dispersion at the Goldstone modes defines the magnon velocity, indicating that the excitation is indeed massless, another characteristics of the spontaneous symmetry breaking. It can also be shown that there might be areas in K-space that have the negative energy excitation as $\mu$ varies. For example, a collinear configuration when $\vec{Q}=\left(0,{2\pi}/\sqrt{3}\right)$ has
\begin{align}
\label{AKBK2}
A_k &= J\left(-1+2\mu+cos{k_x}\right) \notag\\
B_k &= J\left(\mu cos{\frac{k_x}{2}}cos{\frac{\sqrt{3}k_y}{2}}\right)
\end{align}
$A_k^2-B_k^2$ is then expanded around $\vec{k}=\vec{Q}$ as follows
\begin{equation}
\label{expansion1}
A_k^2-B_k^2=J\left\{\left[\left(\mu-1\right)^2-1\right]k_x^2+3\mu^2\left(k_y-\frac{2\pi}{\sqrt{3}}\right)^2\right\}
\end{equation}
from which it is seen that the stable region of the collinear states is $2.0\leq \mu \leq\infty$ in LSWT. (Both $A_k$ and $B_k$ are positive all over the $K$-space). LSWT starts to yield the negative energy excitation in the area surrounding those Goldstone modes when $\mu$ decreases below $2.0$. It is obviously unphysical, meaning that LSWT fails to account for the collinear spin configuration in this parameter regime. On the other hand, a variety of studies,  either theoretical or numerical including our own work, suggest that the collinear configuration is realized when $1.2<\mu\leq\infty$. Therefore it is desirable to investigate it in a way different from LSWT. To this end, we here develop a NLSWT which extends the Bogolyubov transformation to handle negative energy modes. We also introduce a self-consistent iteration scheme.

\section{\label{sec:NLSW}A Novel Nonlinear Spin Wave Theory}

First of all, it is possible that the negative energy excitation exists even for SWT in infinite order if the 
starting spin configuration is unphysical. The emergence of negative energy excitation could also be merely due to an 
insufficient approximation even if that spin configuration physically exists. 
The example here is the negative energy excitation in LSWT for the collinear configuration when $\mu < 2.0$, 
which can be further divided into $\mu \le \mu_0$ and $\mu_0 < \mu < 2.0$ where $\mu_0$ is the physical 
PTP of the collinear configuration. The negative energy excitation might exist in the former regime for infinite order SWT, 
showing a sign of instability of such collinear spin order. While in the latter regime, one should be able to 
get an all positive spectrum in SWT with a sufficient approximation, see below.

Our idea of handling the negative energy modes is simply to proceed like in the Dirac theory of positron \cite{dirac}. 
We apply the particle-hole (anti-particle) transformation, $b \rightarrow b^{\dagger}$, with a boson normal-ordering, 
thereby formally filling up those negative energy modes and regard the filled states as a new physical vacuum.  
Since this formally involves an infinite negative energy,
it is a sort of singular perturbation theory. When combined with mean field treatment of non-linear terms, we arrive at an iterative
scheme of normal ordering and mean field approximations to eliminate negative energy modes and improve on the physical 
vacuum which consists of zero-point motion of all positive energy spectra. Or equivalently, we can make a singular choice, 
\begin{equation}
\label{parity2}
u_k^2-v_k^2=-1
\end{equation}
instead of (\ref{parity1}), for negative energy modes $k$, which leads to the particle-hole transformation, 
\begin{equation}
\label{newcommutation}
\left[b_k,b_k^{\dagger}\right]=\left(u_k^2-v_k^2\right)\left[a_k,a_k^{\dagger}\right]=-1
\end{equation}
, and to a modification of (\ref{LH2}) as
\begin{equation}
\label{LH3}
H_2=\frac{1}{2}\sum_k{\left(\epsilon_k-A_k\right)}+\sum_k{C_k^{*}\epsilon_k b_k^{\dagger}b_k}
\end{equation}
where $C^{*}=1$ for positive energy modes and $-1$ for negative ones. 

In the following two sub-sections we first look at $H_4$ due to its relative simplicity and then $H_3$. After the formulation, we analyze the collinear spin configuration first. It is a self-consistent iterative treatment of $H_4$ alone, since $H_3$ vanishes for the collinear configuration. This is followed by the overall correction combining $H_3$ and $H_4$ together for a general non-collinear spin configuration and anisotropy.

\subsection{\label{SEC:QT}Quartic term}

The quartic term written in the magnon operator reads as
\begin{equation}
\label{QuarticTerm1}
H_4=H_4^1+H_4^2+H_4^3
\end{equation}
where
\begin{align}
\label{QuarticTerm2}
H_4^1 =& \sum_{\delta}{J_{\delta}\sum_{\left(i,j\right)_{\delta}}{cos\left(\vec{Q}\cdot\vec{\delta}\right)a_i^{\dagger}a_ia_j^{\dagger}a_j}} \\
H_4^2 =
& \sum_{\delta}J_{\delta}\sum_{\left(i,j\right)_{\delta}}\frac{1-cos\left(\vec{Q}\cdot\vec{\delta}\right)}{8}\Bigl(a_i^{\dagger}a_j^{\dagger}a_j^{\dagger}a_j\notag\\
& +a_ia_j^{\dagger}a_ja_j+a_i^{\dagger}a_i^{\dagger}a_ia_j^{\dagger}+a_i^{\dagger}a_ia_ia_j\Bigr)
\end{align}
etc. After the Bogolyubov transformation with either regular boson or new boson with a normal ordering, it can be written as
\begin{equation}
\label{QuarticTerm3}
H_4=\delta E_4 +\delta H_2 +\bar{H}_4
\end{equation}
The first term is the Hartree-Fock correction to the ground state energy and the second the magnon self-energy. Since the last term describes two-particle scattering processe, it only yields higher order $1/S$ corrections compared to $\delta E_4$ and $\delta H_2$. So we neglect it here. Before deriving the explicit form of $\delta E_4$ and $\delta H_2$, we apply the Hartree-Fock decoupling to (\ref{QuarticTerm2}). Due to Wick's theorem, one needs to define the following mean-field parameters
\begin{align}
\label{MFP1}
n & \equiv \left\langle a_i^{\dagger}a_i\right\rangle=\sum_k{C_k^{*}v_k^2}\\
m & \equiv \left\langle a_i^{\dagger}a_j\right\rangle=\sum_k{C_k^{*}\gamma_k v_k^2}\\
\Delta & \equiv \left\langle a_ia_j\right\rangle=\sum_k{C_k^{*}\gamma_k u_kv_k}\\
\Omega & \equiv \left\langle a_i^2\right\rangle=\sum_k{C_k^{*} u_kv_k}
\end{align}
To include the effect of anisotropy, one also needs to define $m_{\delta}$ and $\Delta_{\delta}$ respectively as follows
\begin{align}
\label{MFP2}
m_1 & \equiv \left\langle a_i^{\dagger}a_j\right\rangle_{\delta_1}=\sum_k{C_k^{*}cos{k_x} v_k^2}\notag\\
\Delta_2 & \equiv \left\langle a_ia_j\right\rangle_{\delta_2} =\sum_k{C_k^{*}cos\left(\frac{k_x}{2}+\frac{\sqrt{3}k_y}{2}\right) u_kv_k}
\end{align}
etc, where $u_k$ and $v_k$ are defined either to give the regular bosonic commutation or the new commutation for the Bogolyubov transformation depending on the sign of the excitation energy. One should notice that in the extended stable parameter region all the excitation would be positive after the self-consistent iteration converges. Therefore the mixture of bosons and new bosons (they don't interact because they are defined for different wave numbers) will evolve during the iteration. in the converged state, only regular bosons survive. The mean-field parameters are functional of both $\vec{Q}$ and $\vec{k}$. On the other hand, $\gamma_k$ only reflects the average of the nearest neighbors in three orientations of a triangular lattice. It is defined as
\begin{equation}
\label{geometry_average1}
\gamma_k\equiv \frac{1}{6}\sum_{\delta}{e^{i\vec{k}\cdot\vec{\delta}}}=\frac{1}{3}\left(cos{k_x}+2cos{\frac{k_x}{2}}cos{\frac{\sqrt{3}k_y}{2}}\right)
\end{equation}
Now let us look at the following term which is a constant correction to the ground state energy,

\begin{align}
\label{GScorrection1}
\delta E_4=& \sum_{\delta}J_{\delta}\Biggl[ cos\left(\vec{Q}\cdot\vec{\delta}\right)\left(n^2+m_{\delta}^2+\Delta_{\delta}^2\right)\notag\\
& +\frac{1-cos\left(\vec{Q}\cdot\vec{\delta}\right)}{2}\left(2n\Delta_{\delta}+m_{\delta}\Omega\right)\notag\\
& -\frac{1+cos\left(\vec{Q}\cdot\vec{\delta}\right)}{2}\left(2nm_{\delta}+\Delta_{\delta}\Omega\right)\Biggr]
\end{align}
It emerges after the normal-ordering of the quartic term. The left-over two-operator terms are mean-field decoupled, multiplied by the mean-field parameters. For example,

\begin{align}
a_i^{\dagger}a_ia_j^{\dagger}a_j\rightarrow 
& \left\langle a_i^{\dagger}a_i\right\rangle a_j^{\dagger}a_j+\left\langle a_j^{\dagger}a_j\right\rangle a_i^{\dagger}a_i \notag\\
& +\left\langle a_i^{\dagger}a_j^{\dagger}\right\rangle a_ia_j+\left\langle a_ia_j\right\rangle a_i^{\dagger}a_j^{\dagger} \notag\\
& +\left\langle a_i^{\dagger}a_j\right\rangle a_j^{\dagger}a_i+\left\langle a_j^{\dagger}a_i\right\rangle a_i^{\dagger}a_j
\end{align}
The correction to the quantum fluctuation of LSW from the quartic term reads in the $K$-space as follows
\begin{equation}
\label{Qcorrection1}
\delta H_2=\sum_{k}{\delta A_k a_k^{\dagger}a_k-\frac{1}{2}\delta B_k\left(a_k^{\dagger}a_{-k}^{\dagger}+a_ka_{-k}\right)}
\end{equation}
where
\begin{align}
\label{Qcorrection2}
\frac{\delta A_k}{2}=&\left(cos{Q_x}m_1+\frac{1-cos{Q_x}}{4}\Omega-\frac{1+cos{Q_x}}{2}n\right)cos{k_x}\notag\\
&+\mu\left(2f_Qm'+\frac{1-f_Q}{2}\Omega-\left(1+f_Q\right)n\right)f_k\notag \\
&+\frac{1-cos{Q_x}}{2}\Delta_1+\mu\left(1-f_Q\right)\Delta'\notag\\
&-\frac{1+cos{Q_x}}{2}m_1-\mu\left(1+f_Q\right)m'\notag\\
&+\left(cos{Q_x}+2\mu f_Q\right)n\\
\frac{\delta B_k}{2}=&\left(\frac{cos{Q_x}-1}{2}n+\frac{cos{Q_x}+1}{4}\Omega-cos{Q_x}\Delta_1\right)cos{k_x}\notag\\
&+\mu\left[\left(f_Q-1\right)n+\frac{1+f_Q}{2}\Omega-2f_Q\Delta'\right]f_k\notag\\
&+\frac{cos{Q_x}-1}{4}m_1+\frac{cos{Q_x}+1}{4}\Delta_1\notag\\
&+\mu\left(\frac{f_Q-1}{2}m'+\frac{f_Q+1}{2}\Delta'\right)
\end{align}
with the auxiliary mean-filed parameters being defined as follows
\begin{align}
\label{MFP5}
m' &=\sum_{k}{C_k^{*}f_kv_k^2}\notag\\
\Delta' &=\sum_{k}{C_k^{*}f_ku_kv_k}\\
f_k &=cos{\frac{k_x}{2}}cos{\frac{\sqrt{3}k_y}{2}}\notag
\end{align}

\subsection{\label{CT}Cubic term}

The cubic term $H_3$ in (\ref{H3}) reads after the Primakoff transformation as
\begin{align}
\label{CubicTerm1}
& H_3\notag\\
=& \sum_{\left(i,j\right)_{\delta}}{\frac{\sqrt{2S}J_{\delta}}{2}sin\left(\vec{Q}\cdot\vec{\delta}\right)\left[a_i^{\dagger}a_i\left(a_j^{\dagger}+a_j\right)-a_j^{\dagger}a_j\left(a_i^{\dagger}+a_i\right)\right]}
\end{align}
Under the Bogolyubov transformation, it becomes
\begin{align}
\label{CubicTerm2}
H_3=& \sum_{kq}\Bigl[\Gamma_1\left(q,k-q,k\right)b_q^{\dagger}b_{q-k}^{\dagger}b_k\notag\\
& +\Gamma_2\left(q,-k-q,k\right)b_q^{\dagger}b_{-q-k}^{\dagger}b_k^{\dagger}+h.c.\Bigr]
\end{align}
where
\begin{align}
\label{gamma12}
\Gamma_1\left(1,2,3\right)= &\frac{iJ\sqrt{2S}}{2!}[Z_1\left(u_1+v_1\right)\left(u_2u_3+v_2v_3\right)\notag\\
&+Z_2\left(u_2+v_2\right)\left(u_1u_3+v_1v_3\right)\notag\\
&+Z_3\left(u_3+v_3\right)\left(u_1v_2+v_1u_2\right)]\\
\Gamma_2\left(1,2,3\right)= 
&\frac{iJ\sqrt{2S}}{3!}[Z_1\left(u_1+v_1\right)\left(u_2v_3+v_2u_3\right)\notag\\
&+Z_2\left(u_2+v_2\right)\left(u_1v_3+v_1u_3\right)\notag\\
&+Z_3\left(u_3+v_3\right)\left(u_1v_2+v_1u_2\right)]
\end{align}
Above we wrote the expressions symmetric to the wave numbers. $2!$ in $\Gamma_1$ is from the the permutation of $1\leftrightarrow 2$ in $b^{\dagger}_1 b^{\dagger}_2 b_3$ ($b_1 b_2 b^{\dagger}_3$). And $3!$ in $\Gamma_2$ is from the permutation among all three subscripts of $b^{\dagger}_1 b^{\dagger}_2 b^{\dagger}_3$ ($b_1 b_2 b_3$). This symmetric form is somehow more suitable for the numerical integration over the first BZ, especially when there is cancellation of singular terms. Similar to the one from \cite{chernyshev} but (\ref{CubicTerm2}) now includes the anisotropy as follows   
\begin{align}
\label{zkhkgk}
Z\left(Q,k,\mu\right)=& sin{Q_x}sin{k_x}+2\mu\left(h_Qh_k+g_Qg_k\right)\\
h_k=& sin{\frac{k_x}{2}}cos{\frac{\sqrt{3}k_y}{2}}\notag\\
g_k=& cos{\frac{k_x}{2}}sin{\frac{\sqrt{3}k_y}{2}}\notag
\end{align}
The effective Hamitonian reads as
\begin{align}
\label{EffectiveH1}
H_{eff}=H_0+H_3
\end{align}
where $H_0$ either refers to the LSW term $H_2$ or any unperturbed Hamiltonian. If the anisotropy is set to vanish, (\ref{gamma12}) coincides with the ones in \cite{chernyshev, miyake}. Following the standard diagram perturbation method (See Appendix for details), the lowest order contributions are
\begin{align}
\label{selfenergyAB}
{\sum}^a\left(k,\omega\right) &=2\sum_q\frac{\left|\Gamma_1\left(q,k\right)\right|^2}{\omega-\epsilon_q-\epsilon_{k-q}+i0}\\
{\sum}^b\left(k,\omega\right) &=-18\sum_q\frac{\left|\Gamma_2\left(q,k\right)\right|^2}{\omega+\epsilon_q+\epsilon_{k+q}-i0}
\end{align}
to the normal self-energies, and 
\begin{align}
\label{selfenergyCD}
{\sum}^c\left(k,\omega\right)&=6\sum_q\frac{\Gamma_2\left(q,k\right)\Gamma_1^{*}\left(q,-k\right)}{\omega+\epsilon_q+\epsilon_{k+q}-i0}\\
{\sum}^d\left(k,\omega\right)&=-6\sum_q\frac{\Gamma_2\left(q,-k\right)\Gamma_1^{*}\left(q,k\right)}{\omega-\epsilon_q-\epsilon_{k-q}+i0}
\end{align}
to the anomalous self-energies. However the anomalous term does not contribute to the spectrum in the same order $O\left(S^0\right)$ as the normal one's. The new poles of the magnon Green's function will then only reflect the quartic term and the normal self-energies from the cubic term. The real part of the poles determines the renormalized spectrum while the imaginary part determines the damping rate. Explicitly, the new pole is as follows
\begin{align}
\label{finalspectrum1}
\epsilon_k
&=\bar{\epsilon_k}+i\chi_k\notag\\
&=\epsilon_k^0+\sum_q\left(\frac{2\left|\Gamma_1\left(q,k\right)\right|^2}{\epsilon_q^0+\epsilon_{k-q}^0-\epsilon_k^0-i0}+\frac{18\left|\Gamma_2\left(q,k\right)\right|^2}{\epsilon_q^0+\epsilon_{k+q}^0+\epsilon_k^0}\right) 
\end{align}
Since $\Gamma_{1,2}$ are expressed in terms of $u_k$ and $v_k$, it is desirable to write out $\Gamma_{1,2}^2$ explicitly in order to avoid the uncertainty of the sign of $u_k$ and $v_k$. For example,
\begin{align}
\label{GammaSquare1}
\frac{2\Gamma_1^2\left(1,2,3\right)}{SJ^2}=-\sum_{i=1}^{6}T_i
\end{align}
with 
\begin{align}
\label{TI1}
\frac{T_1}{Z_1^2}=&\left(u_1+v_1\right)^2\left(u_2u_3+v_2v_3\right)^2\notag\\
=&\frac{S\left(A_1\right)\left(A_1+B_1\right)}{2\sqrt{\left|A_1^2-B_1^2\right|}}\notag\\
&\left(\frac{S\left(A_2A_3\right)\left(A_2A_3+B_2B_3\right)}{\sqrt{\left|\left(A_2^2-B_2^2\right)\left(A_3^2-B_3^2\right)\right|}}+C_2^{*}C_3^{*}\right)
\end{align}
and
\begin{align}
\label{TI4}
\frac{2T_4}{Z_1Z_2}=&4\left(u_1+v_1\right)\left(u_2u_3+v_2v_3\right)\left(u_2+v_2\right)\left(u_1u_3+v_1v_3\right)\notag\\
=&C_2^{*}C_3^{*}\frac{S\left(A_1\right)\left(A_1+B_1\right)}{\sqrt{\left|A_1^2-B_1^2\right|}}\notag\\
&+C_1^{*}C_3^{*}\frac{S\left(A_2\right)\left(A_2+B_2\right)}{\sqrt{\left|A_2^2-B_2^2\right|}}\notag\\
&+C_1^{*}C_2^{*}\frac{S\left(A_3\right)\left(A_3-B_3\right)}{\sqrt{\left|A_3^2-B_3^2\right|}}
\notag\\
&+\frac{S\left(A_1A_2A_3\right)\left(A_1+B_1\right)\left(A_2+B_2\right)\left(A_3+B_3\right)}{\sqrt{\left|\left(A_1^2-B_1^2\right)\left(A_2^2-B_2^2\right)\left(A_3^2-B_3^2\right)\right|}}
\end{align}
etc. $\Gamma_{2}^2$ is formulated likewise. Note that the sign of the excitation energy of the unperturbed Hamiltonian is included. In the case where the unperturbed spectrum is positive, it has the same expression as after the usual Bogolyubov transformation. However, (\ref{TI4}) should be used whenever one considers NLSWT treatment involving a problematic spectrum from LSWT with a small area of negative energy. More discussion is given in the next two sections.

\section{\label{sec:CollinearConfig} Self-Consistent NSWT Analysis of Collinear Spin Configuration}

The purpose of this paper is to investigate the phase diagram of TAFHM by NLSWT, taking into account the negative energy excitation issue with higher order corrections. For the collinear spin configuration, we need to study the spectrum renormalization due to $H_4$ alone. In the last section, we extend the Bogolyubove transformation to reformulate $\delta H_2$ in (\ref{Qcorrection1}), the contribution from $H_4$ to the quantum fluctuation $H_2$ in LSWT. For the collinear spin configuration, the coefficients in (\ref{Qcorrection1}) are
\begin{align}
\label{AKBK3}
\delta A_k =& \left(2-4\mu\right)n+4\mu\Delta'-2m_1+\left(2m_1-2n\right)cos{k_x} \notag\\
& +\mu\left(2\Omega-4m'\right)f_k\\
\delta B_k =& \left(\Omega-2\Delta_1\right)cos{k_x}+\mu\left(4\Delta'-4n\right)f_k+\Delta_1-2\mu m'
\end{align}
The correction from $H_4$ to the ground state energy, (\ref{GScorrection1}), is written explicitly for the collinear configuration as well
\begin{align}
\label{GScorrection2}
\delta E_4=&\left(1-2\mu\right)\left(n^2+m^2+\Delta^2\right)+2\mu\left(2n\Delta+m\Omega\right)\notag\\
&-\left(2nm+\Delta\Omega\right)
\end{align}

Now we look at the spectrum renormalization due to $\delta H_2$. One can obtain it by
\begin{equation}
\label{Qrenormalization1}
\bar{\epsilon}_k=\epsilon_k+S\left(A_k\right)\frac{A_k\delta A_k-B_k\delta B_k}{\left|\epsilon_k\right|}
\end{equation}
or by
\begin{equation}
\label{Qrenormalization2}
\bar{\epsilon}_k=C_k^{*'}\sqrt{\left|\left(A_K+\delta A_k\right)^2-\left(B_K+\delta B_k\right)^2\right|}
\end{equation}
where $C_k^{*'}$ is determined in the same way as $C_k^{*}$. One immediately sees that when the correction $\delta A_k$ and $\delta B_k$ are small compared with $A_k$ and $B_k$ (\ref{Qrenormalization1}) is almost equivalent to (\ref{Qrenormalization2}) except that (\ref{Qrenormalization1}) has singularity issue at some Goldstone modes where the sign of $\delta B_k/\delta A_k$ is opposite to that of $B_k/A_k$. Physically this singularity will be cancelled either in the summation over the wave numbers in the first BZ or by the other source such as the cubic term in the same order as we will discuss later. However, it is challenging to numerically carry it out. Another drawback of (\ref{Qrenormalization1}) is that the mean-field quantities do not contain the information of the higher order correction. Therefore we prefer to use (\ref{Qrenormalization2}). In fact, the first way can be formulated from the Green's function using $H_2$ as the unperturbed Hamiltonian. The second way can be obtained by absorbing $\delta H_2$ into the unperturbed Hamiltonian. We believe that when the correction is small, these two approximations shall both yield satisfactory results. However so far they are only a half-cooking recipe, because even if they can obtain positive spectrum, neither guarantees the non-recurrance of negative energy problem. The situation would be more sound if a convergence can be provided by a self-consistent iteration. Below we illustrate the idea.

First we start the self-consistent iteration from the ground state of LSWT to get the mean-field parameters. If it has the negative energy excitation for some wave numbers, it is not the lowest energy state but a meta-stable state. When we calculate the mean-field parameters, the new commutation (\ref{parity2}) should be used to collect the contribution of those wave numbers associated with the negative energy excitation. At the end of each iteration one uses the obtained mean-field parameter to update the excitation spectrum. In other words, one re-assesses which excitation is regular boson or the new boson in the first BZ. In the next iteration, the ground state is chosen to be the vacuum of those updated boson and new boson. Still it is a meta-stable state if the negative excitation does not vanish in the whole spectrum. This iteration is carried on until convergence. In the end, there are two possibilities. First, the excitation spectrum becomes all positive, and the ground state has the lowest energy, a physical ground state.  Second, iteration is converged but there still remain some negative energy excitations, signaling the instability of assumed magnetic order.
  
We perform our calculation on a $1800\times 1800$ lattice for which the finite size effect can be neglected. We want to first test the validity of our formulation. To this end, we set $\vec{Q}=\left(0,\frac{2\pi}{\sqrt{3}}\right)$ and $\mu\rightarrow \infty$ to check the well known result of a square lattice with a collinear spin configuration. In this case, there is no negative energy excitation issue even in LSWT. 
\begin{figure}
\begin{center}
\includegraphics[height=2.5in, width=2.5in]{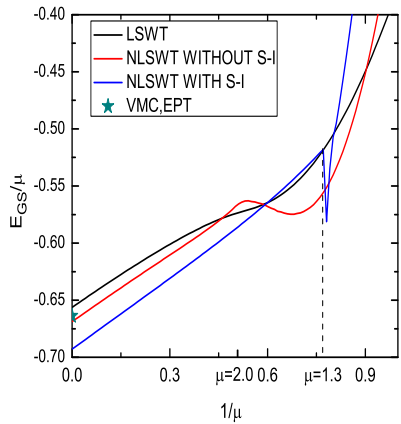}
\caption{\label{CL_COMPARE}LSWT and NLSWT results for the collinear spin configuration on a square lattice by approaching $\mu$ to $\infty$ and setting $\vec{Q}=\left(0,\frac{2\pi}{\sqrt{3}}\right)$.}
\end{center}
\end{figure}
Indeed, in Fig.\ref{CL_COMPARE} one of our NLSWT calculations, which uses (\ref{Qrenormalization1}) and doesn't take a self-consistent iteration process, agrees well with \cite{manousakis}. We also see in Fig\ref{CL_COMPARE} that the energy after the self-consitent iteration converged is obviously further from the believed ground state energy shown by star due to variational Monte Carlo \cite{Huse-Elser,Horsch-Linden,Manousakis1} that is tested by our own nemerical simulation \cite{wang1}. Since self-consistency means to take higher order correlations into account, NLSWT with self-consistent iteration is expected to give a better ground state energy than without self-consistent iteration.  That it is not the case reflects a fact that SWT is not variational and hence an approach to the correct value need not be monotonic.

\begin{figure}
\begin{center}
\begin{tabular}{cc}
(a) $\mu=1.2$ & (b) $\mu=1.3$\\
\includegraphics[height=1.8in, width=1.8in]{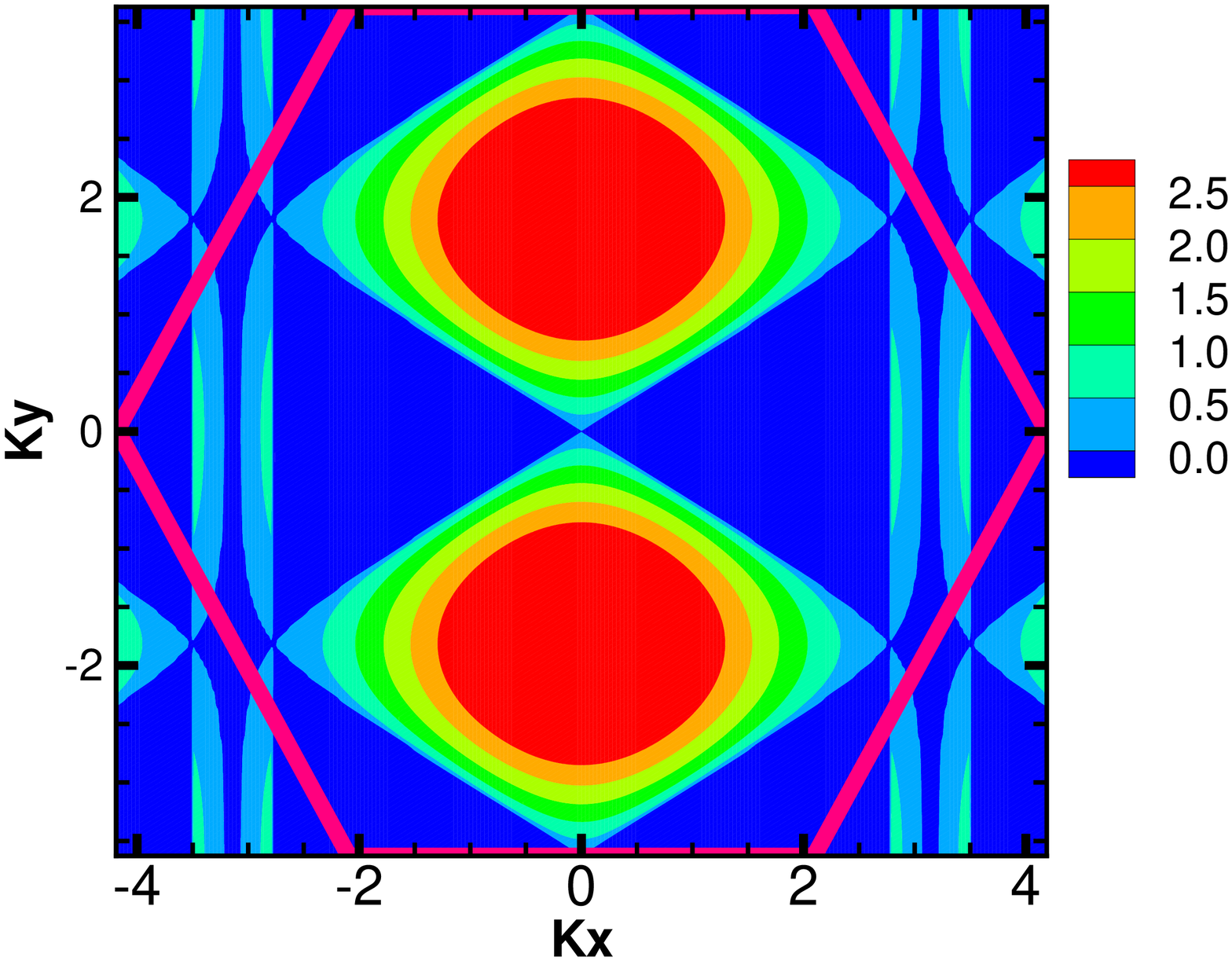} & \includegraphics[height=1.8in, width=1.8in]{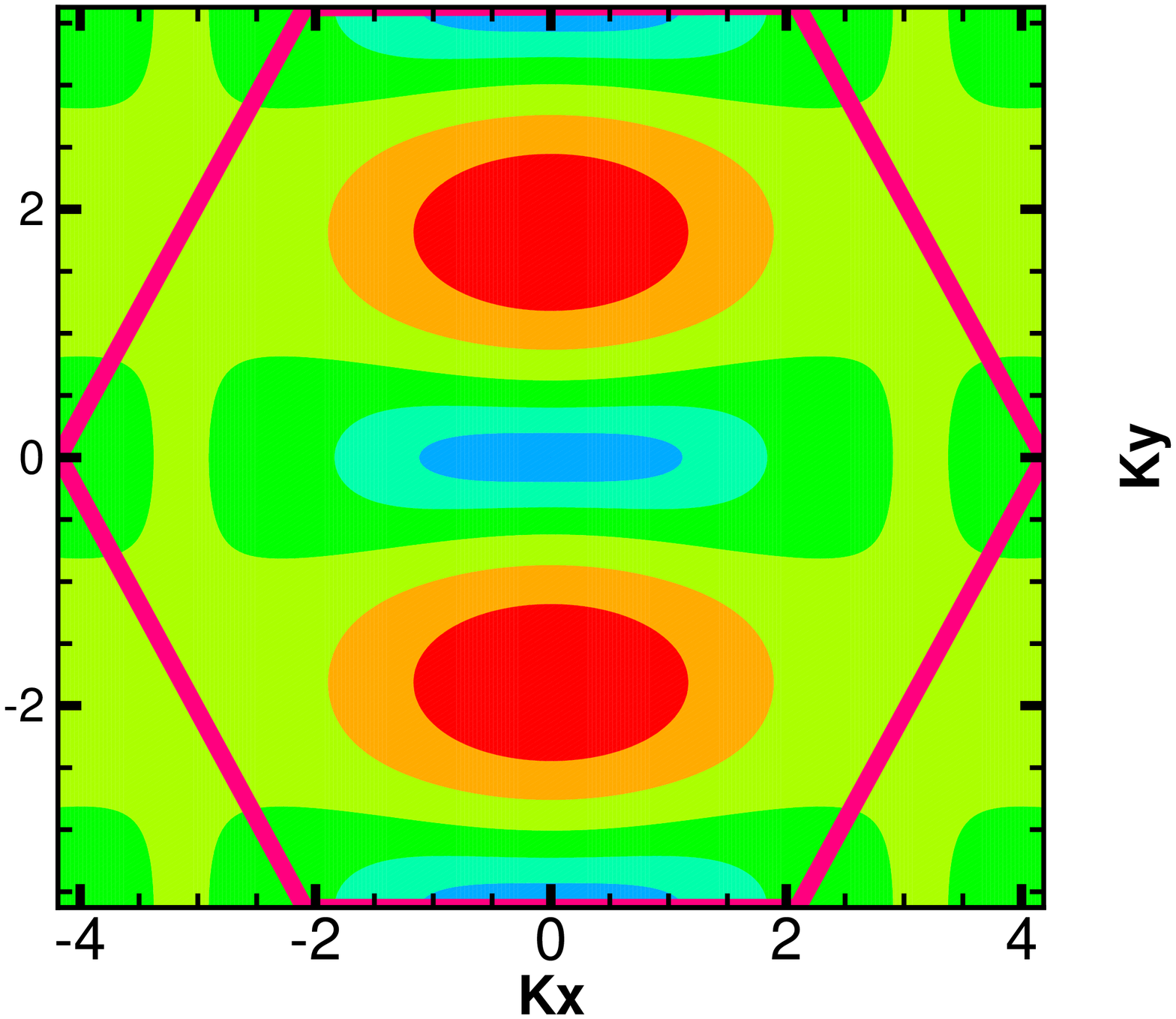}\\
 \end{tabular}
\caption{\label{Fig:CLPHASE}Comparison between the contour plots of the excitation spectrum in the first BZ for (a) $\mu=1.2$ and (b) $1.3$. Both spectra are calculated by NLSWT with a converged self-consistent iteration. The dark blue region in (a) refers to the negative energy excitation. (b) has no negative energy excitation.}
\end{center}
\end{figure}

On the other hand, the self-consistent iteration works better in the intermediate parameter regime in the sense that it improves on the excitation spectrum in $1.3\le\mu\le 2.00$ and that it avoids an artificial oscillation of the NLSWT results without a self-consistent iteration. Furthermore, the discontinuity at $\mu\approx 1.3$ is consistent with the fact that the improvement of spectrum takes effect when $\mu\ge 1.3$ as shown in Fig.\ref{Fig:CLPHASE}. That is to say, it provides a clear and consistent sign that there is a phase transition from the collinear spin configuration to another configuration when $\mu$ decreases below $1.3$. 

\begin{figure}
\begin{center}
\begin{tabular}{cc}
(a) Around NLSWT PTP & (b) Around LSWT PTP\\
\includegraphics[height=1.8in, width=1.8in]{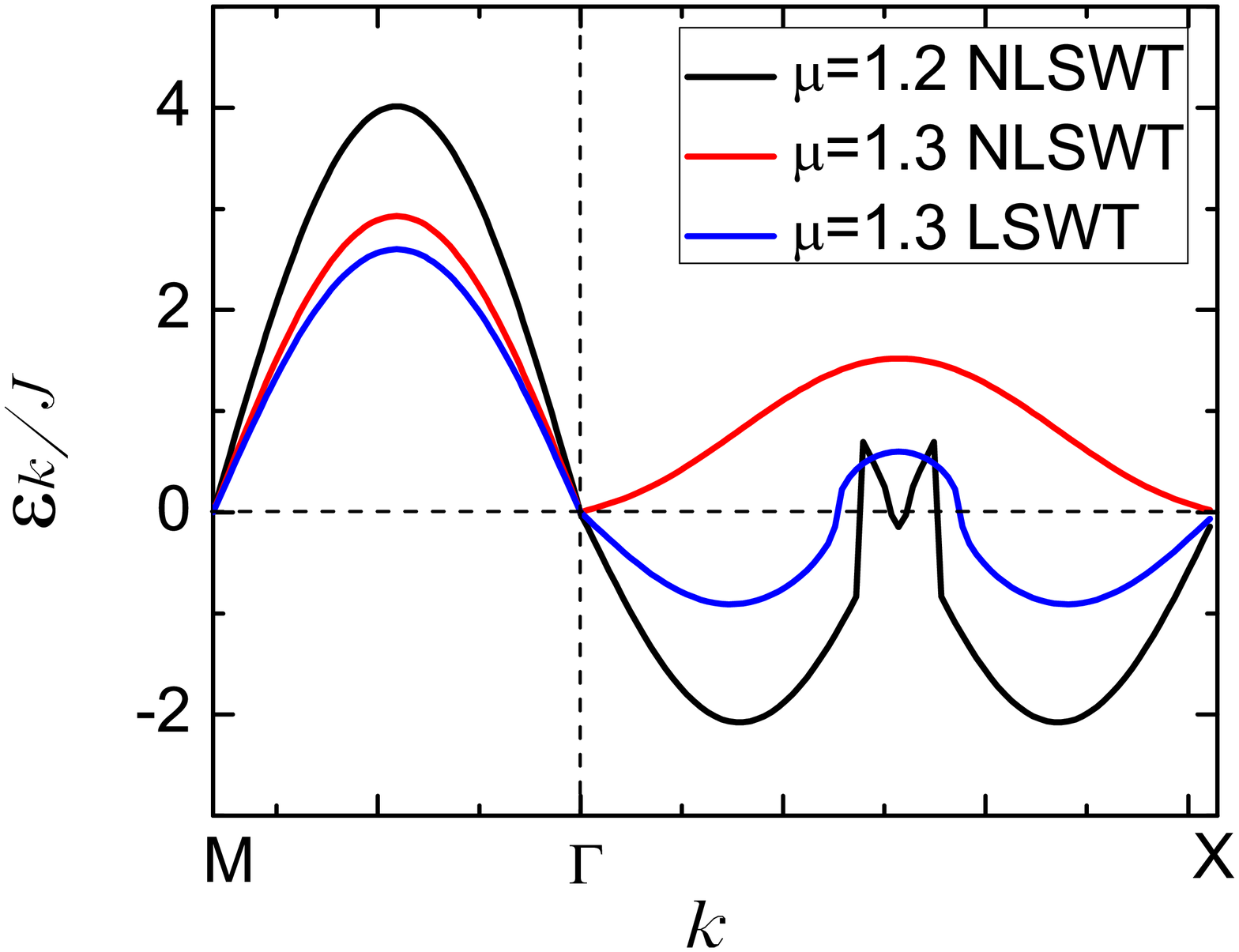} & \includegraphics[height=1.8in, width=1.8in]{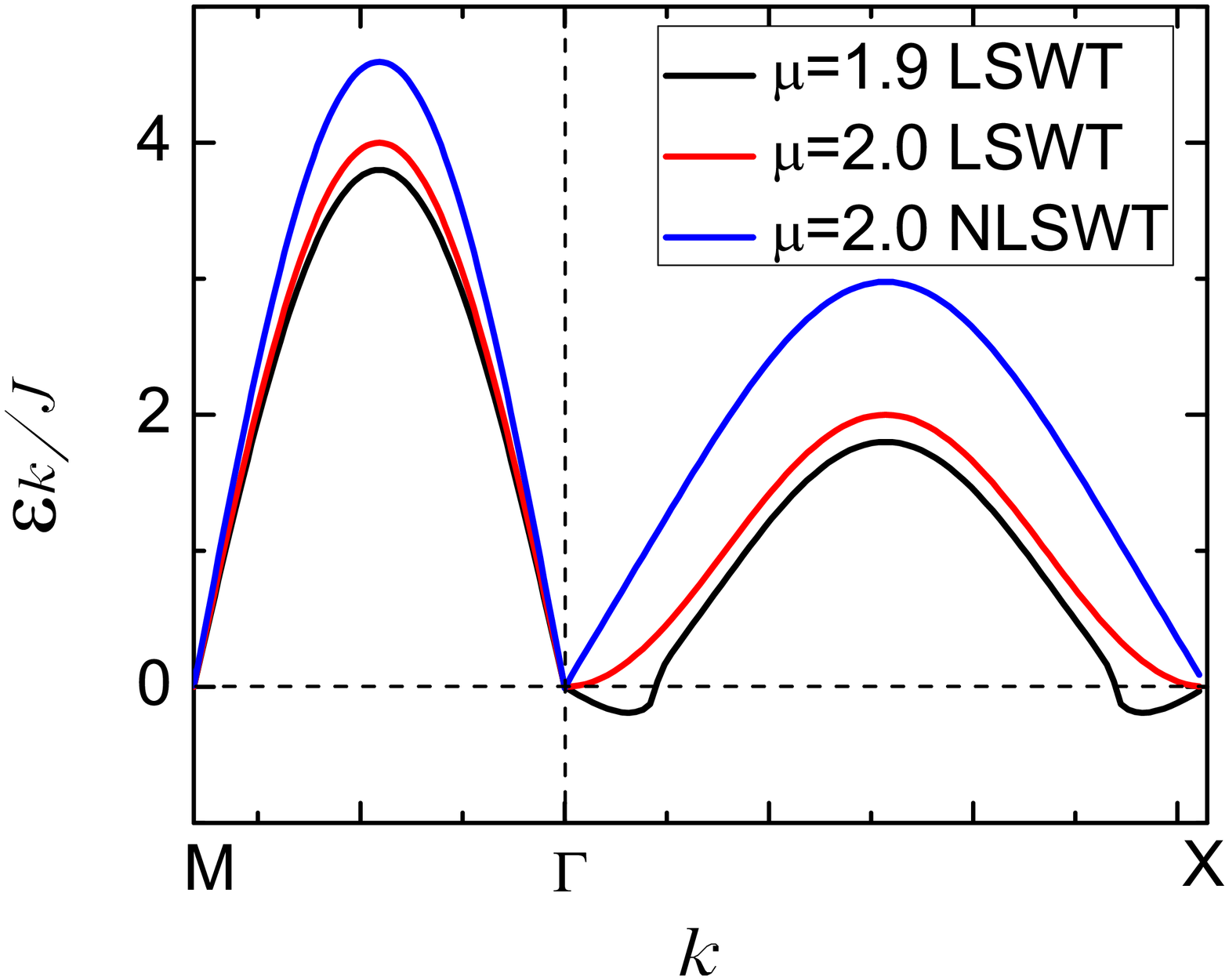}\\
 \end{tabular}
\caption{\label{Fig:around_PTP}Comparison between the spectra of LSWT and NLSWT around the phase transition point (PTP) at (a) $\mu=1.3$ for NLSWT and (b) $2.0$ for LSWT. }
\end{center}
\end{figure}

It is known that PTP of collinear configuration in LSWT is $\mu= 2.0$. In contrast, it is $\mu\approx 1.3$ in our NLSWT. So we compare the excitation spectrum between LSWT and our NLSWT around these two points. Fig.\ref{Fig:around_PTP} shows the comparison (a) around the PTP of NLSWT and (b) around the PTP of LSWT. Besides the obvious occurance of negative energy excitation when $\mu$ decreases below their PTPs, the softening of the spin wave at $\Gamma$-point deserves an emphasis. The strongest softening happens when the sound velocity is calculated along the direction from $\Gamma$-point to X-point. For example, it is zero at $\mu=2.0$ in LSWT, implying a breakdown.

\begin{figure}
\begin{center}
\includegraphics[height=2.5in, width=2.5in]{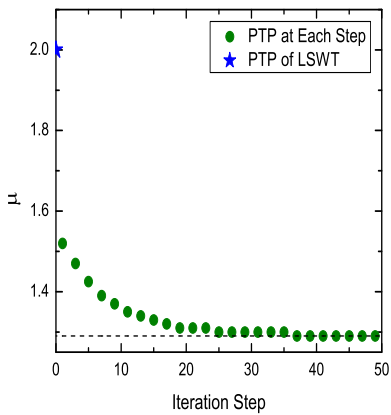}
\caption{\label{Fig:PTP}The star (blue online) show PTP of LSWT. It moves to $\mu=1.52$ immediately after one iteration. Eventually it converges to $\mu\approx 1.3$.}

\end{center}
\end{figure}

It is interesting to see how PTP evolves during the iteration. We change $\mu$ in the range of $1.2\le \mu \le 2.0$ and for each value investigate the excitation spectrum at every step of iteration. We are then able to determine PTP vs iteration steps (IS) shown in Fig.\ref{Fig:PTP}. One sees that the PTP of the collinear spin configuration approaches quickly to the convergence right after a few iteration. It is an important feature, not only because it reveals that the new commutation relation renormalizes the spectrum systematically (In fact, without a correct treatment of the negative energy excitation we found no way to renormalize the spectrum to the physically meaningful one while the correct symmetry, i.e., the Goldstone mode, is intact), but also because this nice tendency yields a useful 
information about the phase transition even when it does not converge.

\section{\label{sec:NoncollinearConfig} NLSWT Analysis of Non-Collinear Spin Configuration}

We also want to test our method for non-collinear spin configurations and anisotropy by applying it to the limiting case which can be readily compared with the existing calculations. We discuss in Sec.\ref{SEC:Commensurate} the commensurate spiral order, $\vec{Q}=\left(4\pi/3,0\right)$ and $\mu=1$. It is a $120^o$ three sub-lattice spin configuration on the isotropic lattice. In Sec.\ref{SEC:INCommensurate} we extend the discussion to the incommensurate spiral order. 

\subsection{\label{SEC:Commensurate}$120^o$ Commensurate Spiral Order}
\begin{figure}
\begin{center}
\begin{tabular}{cc}
(a) & (b) \\
\includegraphics[height=1.8in, width=1.8in]{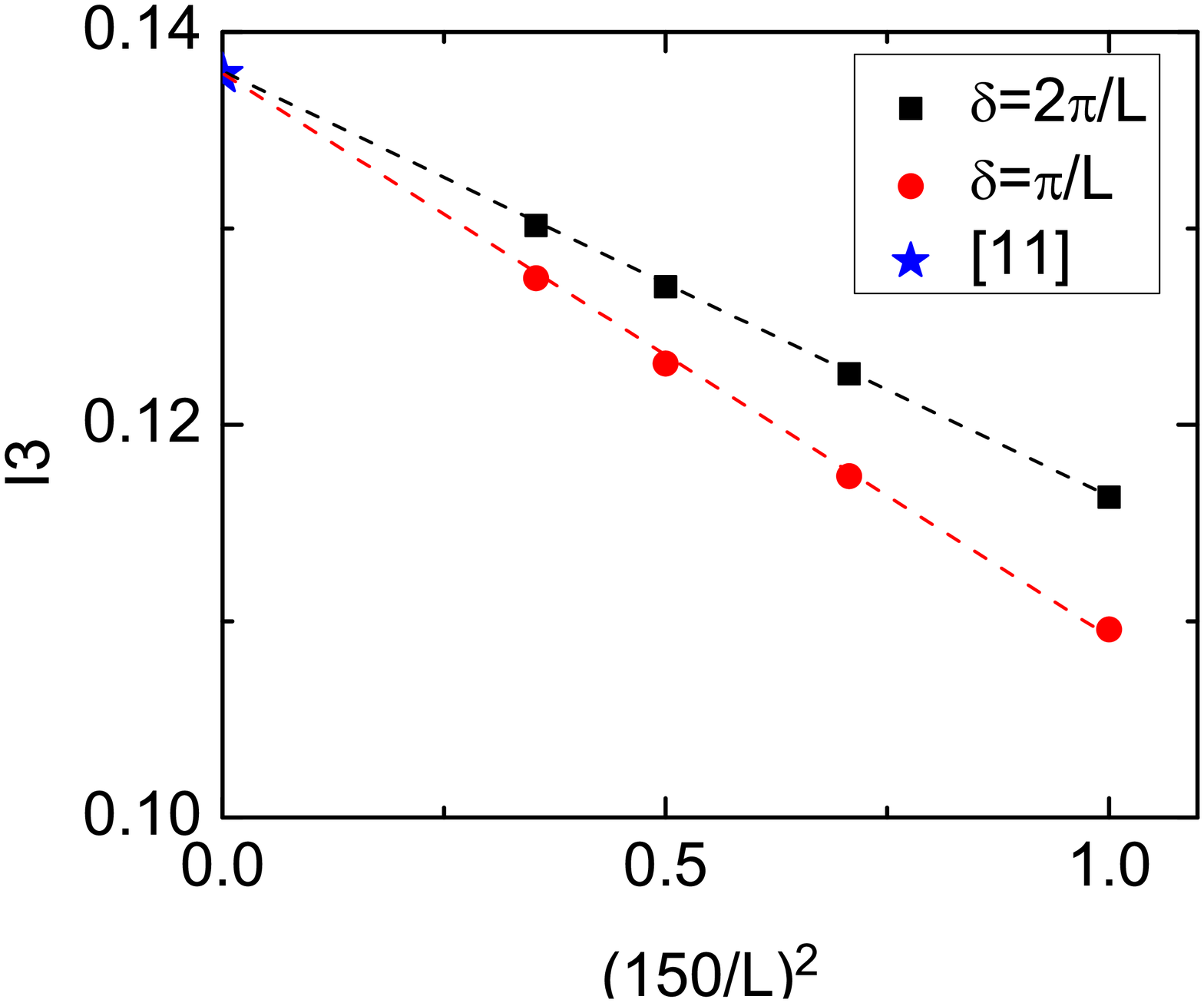} & \includegraphics[height=1.8in, width=1.8in]{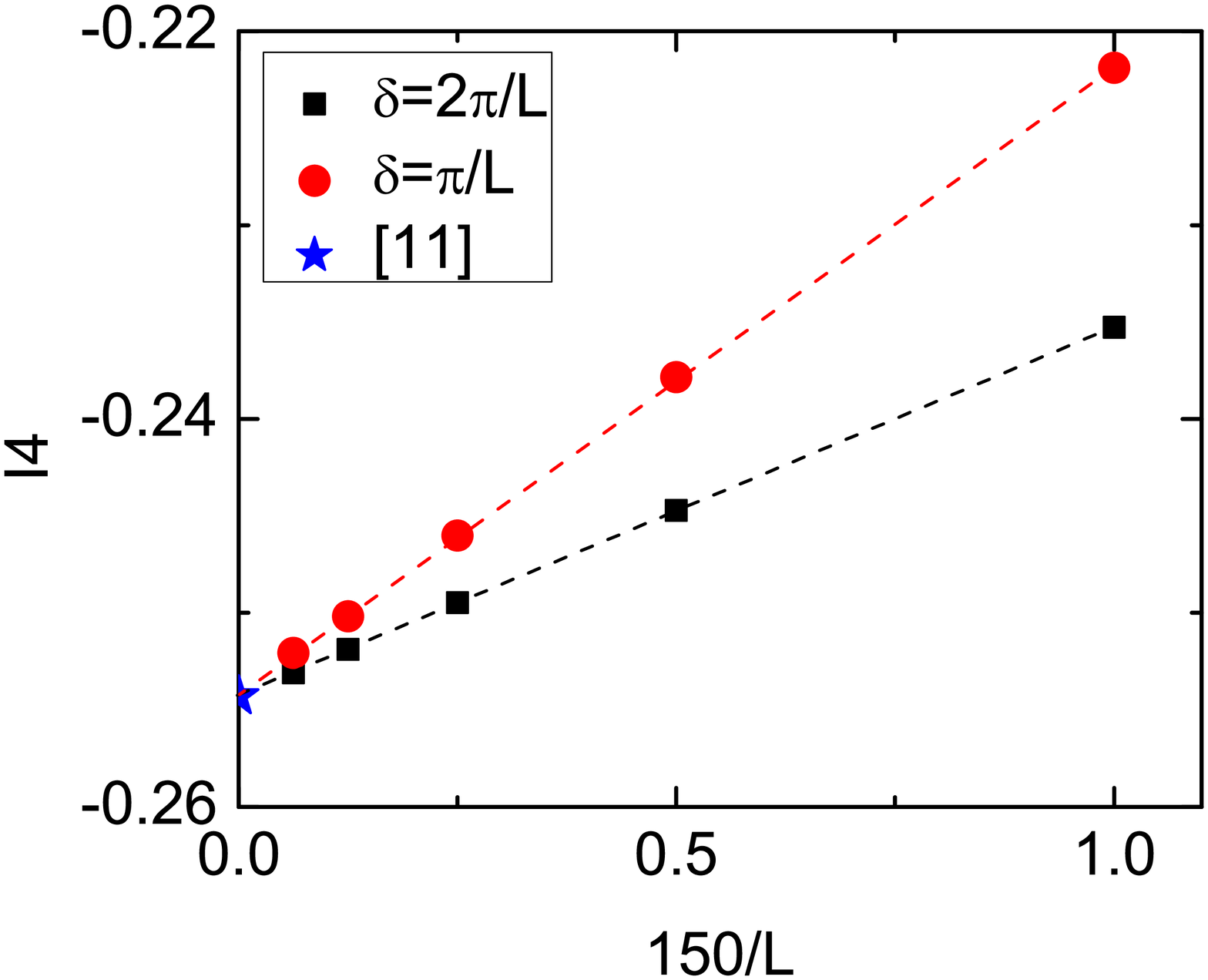}\\
 \end{tabular}
\caption{\label{Fig:scaling1} The finite-size scaling of the contribution to the ground state energy from (a) CT and (b) QT. This scaling is for the $120^o$ spiral configuration on an isotropic lattice} 
\end{center}
\end{figure}

Note that the contributions from the cubic term (CT) and from the quartic term (QT) both have the singularity at $\vec{k}=\pm \vec{Q}$. From Fig.\ref{Fig:scaling1} one sees that their singularities exactly cancel out only at the infinite limit. Since the numerical calculation can be done only on a finite lattice, a finite-size scaling is necessary to extrapolate the precise corrections. For example we show two quantities that are related to the contribution to GSE, $I3$ from CT in (a) of Fig.\ref{Fig:scaling1} and $I4$ from QT in (b) of the same figure. Particularly for an isotropic lattice $I4\equiv -8\delta E_4 /3J$ with $\delta E_4$ being defined in (\ref{GScorrection1}) and $I3\equiv -4\delta E_3/3J$ with $\delta E_3$ being the contribution to GSE from CT. $I3$ reads as follows
\begin{align}
\label{EQ:I3}
I3=8\frac{\left|\Gamma_2\left(q,k\right)\right|^2}{\epsilon_q^0+\epsilon_{k+q}^0+\epsilon_k^0}
\end{align}
One sees from Fig.\ref{Fig:scaling1} that $I3$ converges much slower than $I4$. Note that our direct numerial simulation 
is much heavier compared to the finite-size extrapolation using clusters with different aspect ratios\cite{white}, and gives a
clear convergence as seen in Fig.\ref{Fig:scaling1}. The extrapolated values of $I3$ and $I4$ agree with \cite{chernyshev, miyake}
denoted by blue stars. In addition, it is well known that in a direct numerical simulation a small number $\delta$, comparable with the increment of numerical integration in the $K$ space, should be inserted in the denominator of any fractional number. It is to prevent the overflow in the numerical simulation. Therefore finite-size scaling using different $\delta$ is presented in the figure. Smaller $\delta$ gives a faster convergence. The standard choice $\delta=\pi/L$ is also good as seen in the figure.       

\begin{figure}
\begin{center}
\includegraphics[height=2.5in, width=2.5in]{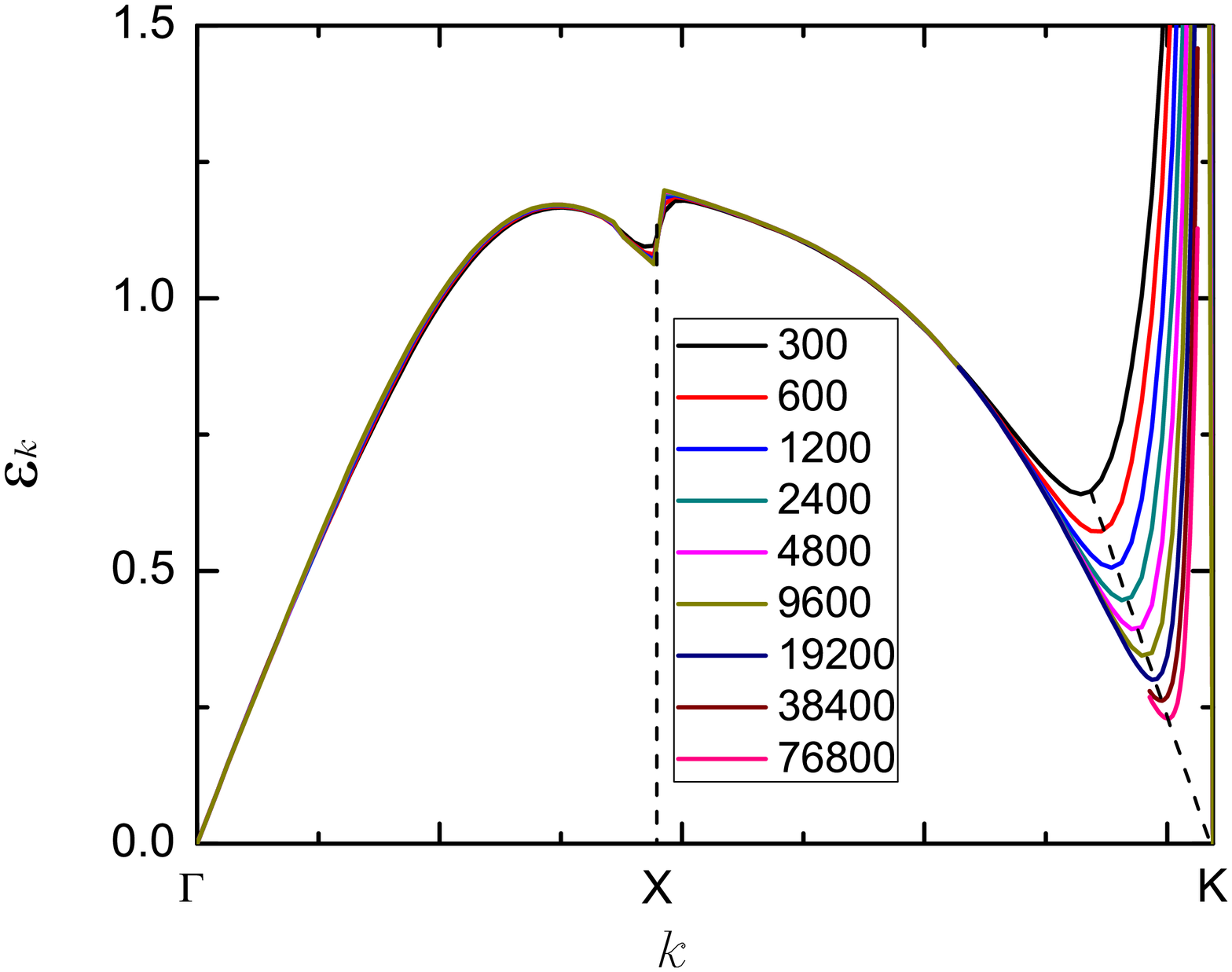} 
\caption{\label{Fig:scaling2}The finite-size scaling of the spectrum for the $120^o$ spiral configuration on an isotropic lattice. The slice from $\Gamma$ point, $\vec{k}=\left(0,0\right)$ to $K$ point, $\vec{k}=\vec{Q}=\left(4\pi/3,0\right)$, is shown.}
\end{center}
\end{figure}

We then use $\delta=\pi/L$ to perform the numerical simulation for the spectrum. A finite-size scaling is shown in Fig.\ref{Fig:scaling2}. It is clear that the singularities from CT and QT at $K$-point indeed exactly cancel out at the thermodynamic limit but it scales very slowly. The spectra from top to bottom in the figure account for different lattice sizes, from $L=300$ to $76800$. The extrapolated spectrum agrees with \cite{chernyshev}.

It is also notable that there exists discontinuity in the spectrum, shown as $X$-point in Fig.\ref{Fig:scaling1}. It could be cured by some techniques. For example the damping can be absorbed into the Dyson equation to further renormalize the spectrum like in the off-shell approximation \cite{chernyshev}. However we don't go in this direction for two reasons. First, the $120^o$ spiral order when $\mu=1.0$ does not have negative energy excitation despite the discontinuity. Second, those further treatments like off-shell approximation are unlikely to cure the negative energy excitation when $\mu$ varies away from $1.0$ while the configuration is fixed. Instead, in the next section we discuss the instability of the incommensurate spiral order in NLSWT, which is originally stable in LSWT at specific values of $\mu$. 

\subsection{\label{SEC:INCommensurate}Incommensurate Sprial Order}
There exist incommensurate spiral states in LSWT when $\mu$ varies away from $1.0$. The relationship between $\mu$ and the ordering wave number is:  $\vec{Q}=\left(2cos^{-1}\left(-\mu/2\right),0\right)$ when $0.27<\mu<1.0$ and $\vec{Q}=\left(2\pi-2cos^{-1}\left(-\mu/2\right),2\pi/\sqrt{3}\right)$ when $1.0<\mu<2.0$. Note that the splitting into branches is just due to the folding into the first BZ of an originally continuous line. A simple calculation shows that this relationship is identical to the one in the classical picture except that it extends into $0<\mu<0.27$. The difference can be explained if the reduction, the difference between the magnetization and the classical value ($1/2$ for spin one half), is considered. It is found that the reduction exceeds the classical value when $0<\mu<0.27$ in LSWT, indicating its breakdown in this regime and that the state is disordered. 

\begin{figure}
\begin{center}
\begin{tabular}{cc}
(a) & (b) \\
\includegraphics[height=1.8in, width=1.8in]{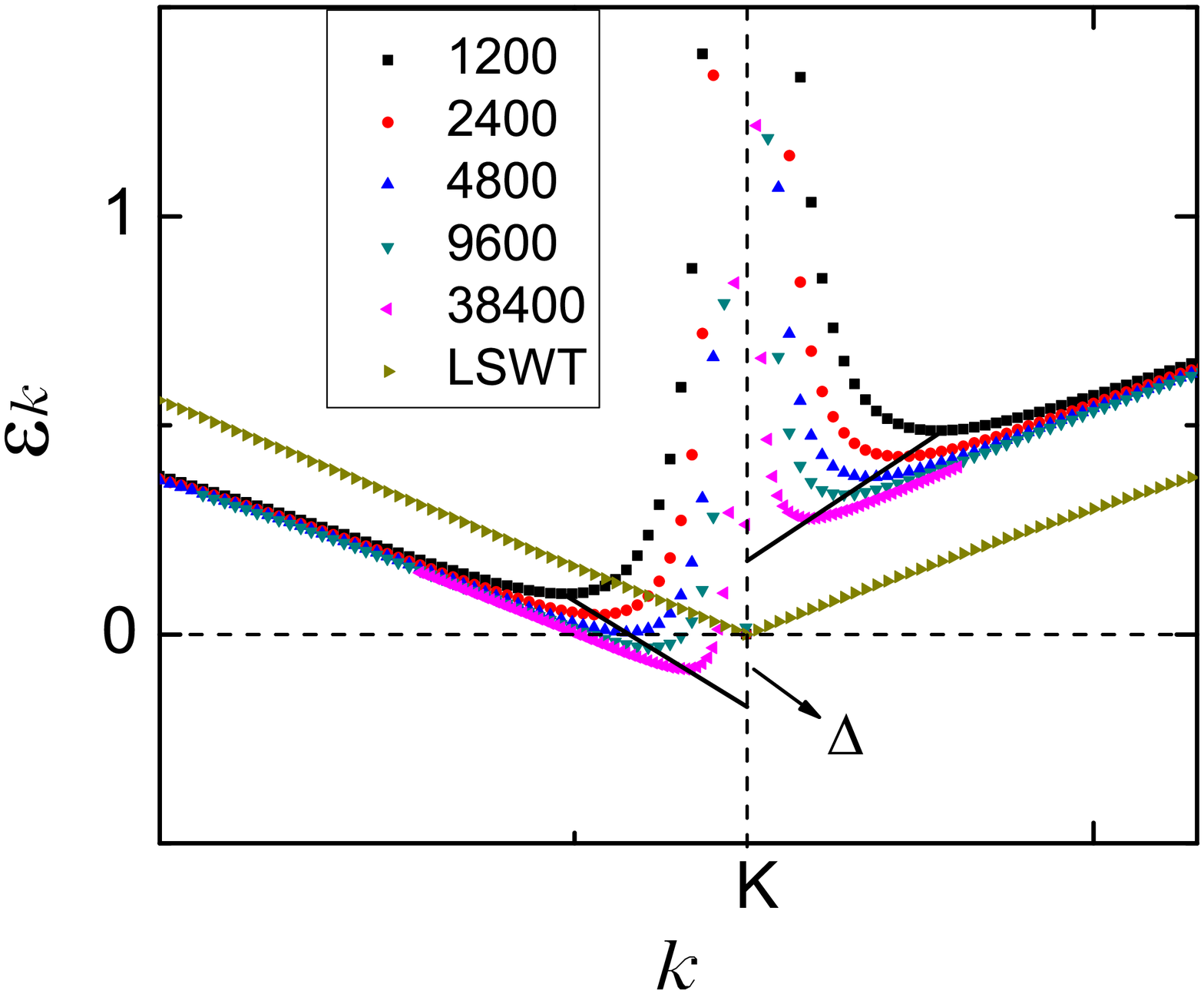} & \includegraphics[height=1.8in, width=1.8in]{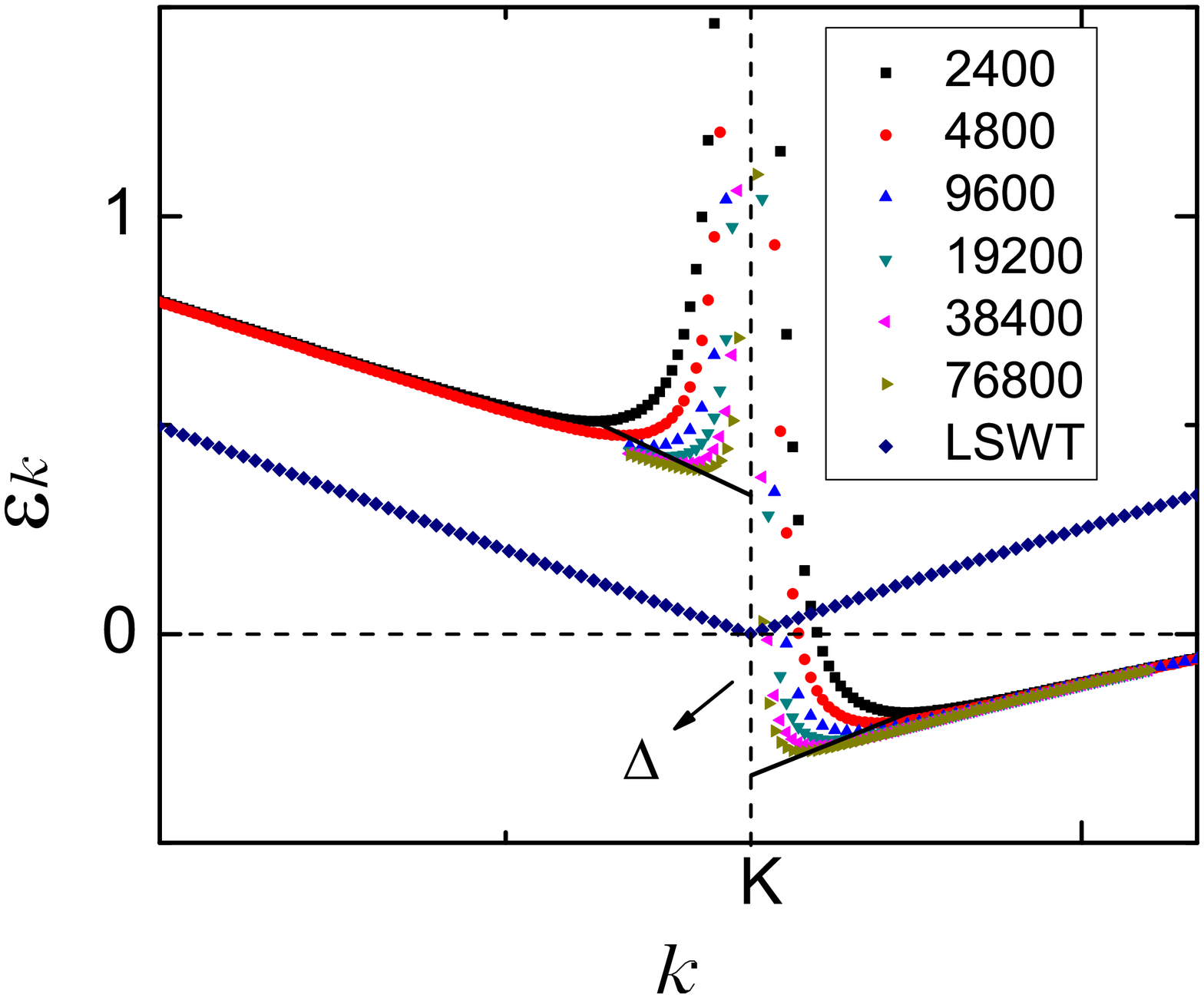}\\
 \end{tabular}
\caption{\label{Fig:scaling3} The finite-size scaling of the spectrum for incommensurate spiral configurations. A slice is shown, from $\Gamma$ point to $K$-point of (a) $\left(7\pi/6,0\right)$ and (b) $\left(\pi/2,2\pi/\sqrt{3}\right)$} 
\end{center}
\end{figure}

We study these incommensurate spiral configurations by NLSWT in this session. A few wave numbers are chosen to carry out the numerical simulation. The corresponding value of $\mu$ resides on both sides of $1.0$. And only two typical spectra are plotted. They are (a) of Fig.\ref{Fig:scaling3} when $\vec{Q}=\left(7\pi/6,0\right)$ and (b) $\left(\pi/2,2\pi/\sqrt{3}\right)$. The spectrum is plotted along the same slice as in Fig.\ref{Fig:scaling2}, with only the vicinity of $K$-point, one of the Goldstone modes, being shown.

\begin{figure}
\begin{center}
\includegraphics[height=2.5in, width=2.5in]{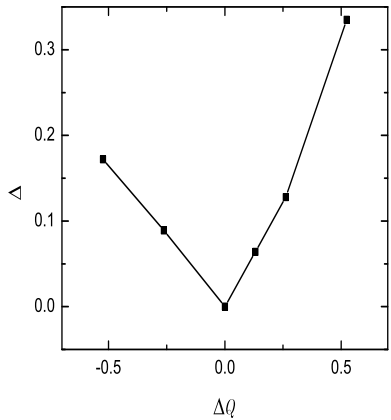} 
\caption{\label{Fig:depth}The finite-size scaling of the spectrum for a $120^o$ spiral configuration on an isotropic lattice. The slice from $\Gamma$ point, $\vec{k}=\left(0,0\right)$ to $K$ point, $\vec{k}=\vec{Q}=\left(4\pi/3,0\right)$, is shown.}
\end{center}
\end{figure}

The solid lines show the extrapolated spectrum by the finite-size scaling. They exhibit both the negative energy excitation and a discontinuity at $K$-point. The relationship between the distance from the commensurate wave number, $\Delta Q\equiv Q_x-4\pi/3$ with $Q_x$ being the unfolded incommensurate wave number, and $\Delta$, the depth of the negative energy, i.e., half of the kink since the discontinuity is symmetric about $0$, is plotted in Fig.\ref{Fig:depth}. It is linear near the commensurate spiral wave number. Although only a few wave numbers are checked, it seems in Fig.\ref{Fig:depth} that all the incommensurate wave numbers are corresponding to spectra with negative energy excitation and discontinuity at $K$-point. Namely, the incommensurate spiral spin configuration is unstable in the NLSWT.

\subsection{\label{SEC:proposal}Summary and Proposal of a Self-Consistent Iteration}
From the last two sections we conclude that in this NLSWT only the commensurate $120^o$ spiral order is stable among all the spiral configurations. But it would be interesting to see if special treatments like the off-shell approximation can cure the discontinuity at the $K$-point, say, at least for some incommensurate spiral wave number. If so, the depth of the negative energy will then disappear because it is tightly associated with the discontinuity. However, this perspective is essentially different from what off-shell approximation does for the commensurate spiral state where the discontinuity at $X$-point is smoothed by it while the spectrum at $K$-point is not altered.

\section{\label{sec:conclusion} Conclusion and Discussion}
We have extended the Bogolyubov transformation by considering a new commutation relation for the bosonic quasi-particle which has negative energy. This new transformation, equivalent to a boson normal ordering, bridges a self-consistent iteration from a spectrum with 
areas of negative energy excitation in LSWT to an expected positive energy spectrum in NLSWT. The new stable regime $1.3<\mu<\infty$ for the collinear spin configuration on a spin $1/2$ TAFHM is comparable with the ones by other methods \cite{MQWeng, bernu, yunoki, wang1}. Considering the clear physical picture of both the ground state and the excitation spectrum provided by SWT, such improvement of the approximation precision over the previous SWT studies \cite{capriotti,trumper,merino} stimulates our interest to apply the same scheme to non-collinear spin configurations. 
Indeed our study shows that among spiral configurations, only the $120^o$ commensurate sprial is stable at $\mu=1.0$ while all incommensurate ones are unstable. This findling has two-fold meanings. First, most incommensurate sprial states are not seen by 
previous studies, which coincides with this finding. Second, however, certain incommensurate spiral state is seen. For instance, $\vec{Q}=\left(\pi/2,2\pi/\sqrt{3}\right)$, namely a four sub-lattice spiral state has been seen in a very narrow window around $\mu=1.18$ by our recent Eangtanglement Perturbation Theory study \cite{wang1}. It implies that due to the approximate nature of SWT, especially 
when the spin is as small as $1/2$, NLSWT still misses certain details in the phase diagram. Moreover, our more elaborate 
self-consistent iteration study for a non-collinear spin configuration does not give a satisfactory result of an expected broadening 
of the stable region for $120^o$ spiral state. 
A further development of NLSWT along this line shall be highly desirable.
\section{\label{sec:acknownledgement} Acknownledgement}
Lihua Wang would like to acknowledge the fruitful discussion with Dr. Kazuo Ueda. The numerical calculation of this work utilized the RIKEN Integrated Cluster of Clusters at Advanced Center for Computing and Communication, RIKEN.



\appendix 
\section{\label{sec:appendix} Self-Energy}

In this appendix, we show the formulation of the self-energy in the lowest order for CT, for instance, ${\sum}^a\left(k,\omega\right)$. The green function is defined as follows
\begin{align}
\label{greenfunction1}
G\left(k,\tau\right)=-\left\langle T_{\tau}\tilde{b}_k\left(\tau\right)b_k^{\dagger}\left(0\right)\right\rangle=-\left\langle T_{\tau}U\left(\beta\right)b_k\left(\tau\right)b_k^{\dagger}\left(0\right)\right\rangle
\end{align}
with
\begin{align}
\label{smatrix1}
U\left(\beta\right)=\sum_{n=0}^{\infty}\left(\frac{-1}{h}\right)^n\frac{1}{n!}\int_0^{\beta}d\tau_1\cdots d \tau_n\left\langle T_{\tau}H'\left(\tau_1\right)\cdots H'\left(\tau_n\right)\right\rangle
\end{align}
The perturbative Hamiltonian is $H'=H^4+H^3$. One immediately sees that $H^4$, if one choose not to absorb it to $H^0$, contributes in (\ref{smatrix1}) as the first order, it will give $\tilde{\epsilon}_k$, just a HF approximated term as we discussed before. Now we focus on the self-energy due to $H^3$, which arises from the expansion of (\ref{smatrix1}) in the second order:
\begin{align}
\label{U2}
U^{\left(2\right)}\left(\beta\right)=\frac{1}{h^2}\frac{1}{2!}\int_0^{\beta}d\tau_1 d\tau_2 H^3\left(\tau_1\right)H^3\left(\tau_2\right)
\end{align}
Therefore
\begin{align}
\label{greenfunction2}
G^{\left(2\right)}\left(k,\tau\right)=-\frac{1}{h^2}\frac{1}{2!}\int_0^{\beta}d\tau_1 d\tau_2 \left\langle H^3\left(\tau_1\right)H^3\left(\tau_2\right)b_k\left(\tau\right)b_k^{\dagger}\left(0\right)\right\rangle
\end{align}
Using (\ref{CubicTerm2}) we have
\begin{widetext}
\begin{align}
\label{green3}
H^3\left(\tau_1\right)H^3\left(\tau_2\right)
= \sum_{q_1 k_1}\Gamma_1\left(q_1,k_1-q_1,k_1\right)\sum_{q_2 k_2} \Gamma_1^{*}\left(q_2,k_2-q_2,k_2\right) b_{q_1}^{\dagger}\left(\tau_1\right)b_{q_1-k_1}^{\dagger}\left(\tau_1\right)b_{k_1}\left(\tau_1\right)b_{q_2}\left(\tau_2\right)b_{q_2-k_2}\left(\tau_2\right)b_{k_2}^{\dagger}\left(\tau_2\right)+\cdots
\end{align}
\end{widetext}
The first term above will give rise to ${\sum}^a\left(k,\omega\right)$, Now we show the formulation step by step. Due to Wick's theorem, one of terms in the form of operator-pair to be summed in (\ref{greenfunction2}) is
\begin{widetext}
\begin{align}
\label{wick1}
& \left\langle T_{\tau}b^{\dagger}_{q_1}\left(\tau_1\right)b_{q_2}\left(\tau_2\right)\right\rangle  \left\langle T_{\tau}b^{\dagger}_{k_1-q_1}\left(\tau_1\right)b_{k_2-q_2}\left(\tau_2\right)\right\rangle \left\langle T_{\tau}b_{k_1}\left(\tau_1\right)b_{k}^{\dagger}\left(0\right)\right\rangle  \left\langle T_{\tau}b^{\dagger}_{k_2}\left(\tau_2\right)b_{k}\left(\tau\right)\right\rangle \notag\\
= & \left\langle T_{\tau}b_{q_2}\left(\tau_2\right)b^{\dagger}_{q_1}\left(\tau_1\right)\right\rangle \delta_{q_1,q_2}\left\langle T_{\tau}b_{k_2-q_2}\left(\tau_2\right)b^{\dagger}_{k_1-q_1}\left(\tau_1\right)\right\rangle \delta_{k_1-q_1,k_2-q_2}\left\langle T_{\tau}b_{k_1}\left(\tau_1\right)b^{\dagger}_{k}\left(0\right)\right\rangle \delta_{k_1,k}\left\langle T_{\tau}b_{k}\left(\tau\right)b^{\dagger}_{k_2}\left(\tau_2\right)\right\rangle \delta_{k,k_2}\notag\\
= & G_0\left(q_1,\tau_2-\tau_1\right)\delta_{q_1,q_2} G_0\left(k_1-q_1,\tau_2-\tau_1\right)\delta_{k_1-q_1,k_2-q_2} G_0\left(k,\tau_1\right)\delta_{k_1,k} G_0\left(k,\tau-\tau_2\right)\delta_{k,k_2}
\end{align}
\end{widetext}
Where $G_0$ is the unperturbed Green function,
\begin{equation}
\label{greenfunction4}
G_0\left(k,\tau\right)=-\left\langle T_{\tau} b_k\left(\tau\right) b_k^{\dagger}\left(0 \right) \right\rangle
\end{equation}
Substituting (\ref{greenfunction4}) into (\ref{wick1}) and using the kronig-delta function one has
\begin{align}
\label{haha1}
G_1^{\left(2\right)}\left(k,\tau\right)= & -\frac{1}{2h^2}\int_0^{\beta}d\tau_1d\tau_2 \sum_q \left|\Gamma_1\left(q,k\right)\right|^2 G_0\left(q,\tau_2-\tau_1\right)\notag\\
& G_0\left(k-q,\tau_2-\tau_1\right) G_0\left(k,\tau-\tau_2\right) G_0\left(k,\tau_1\right)
\end{align}
The first subscript stands for the lowest order while the second for only part of terms. After the Frourier transformation, (\ref{haha1}) reads as
\begin{align}
\label{haha2}
& G_1^{\left(2\right)}\left(k,i\omega\right)\notag\\
= & \frac{1}{2\beta}\sum_q \sum_{\omega'}\left|\Gamma_1\left(q,k\right)\right|^2 G_0\left(q,i\omega'\right)  G_0\left(k,i\omega\right)\notag\\
& G_0\left(k,i\omega\right)G_0\left(k-q,i\omega-i\omega'\right) \notag\\
= & G_0\left(k,i\omega\right){\sum}^a_1\left(k,\omega\right)G_0\left(k,i\omega\right) 
\end{align}
Thus one gets the implicit expression of part of the first normal self-energy  
\begin{align}
\label{haha3}
{\sum}^a_1\left(k,\omega\right)= & \frac{1}{2\beta}\sum_{\omega'}\sum_q \left|\Gamma_1\left(q,k\right)\right|^2 \notag\\
& G_0\left(q,i\omega'\right) G_0\left(k-q,i\omega-i\omega'\right)
\end{align}
If one needs its explicit expression, the unperturbed Green function above must be integrated out with respect to $q$ and $\omega'$. Take the definition\cite{chernyshev}
\begin{align}
\label{haha4}
G_0\left(k,i\omega\right) =& \frac{1}{i\omega-\omega_k} 
\end{align}
And note that the boson distribution function is
\begin{align}
\label{bosondistribution}
n_B\left(z\right)=\frac{1}{e^{\beta z}-1}
\end{align}
Its poles are the Matsubara frequencies for boson. One then transforms the frequency summation to an integral
\begin{align}
\label{contourintegral}
\frac{1}{\beta}\sum_{i\omega} \rightarrow & \frac{1}{2\pi i}\oint_c dz n_B\left(z\right)
\end{align}
Explicitly, one defines a contour integral as follows
\begin{align}
\label{final1}
& I\notag\\
=& \frac{1}{2}\sum_{q} \left|\Gamma_1\left(q,k\right)\right|^2 \left[\frac{1}{2\pi i}\oint_c dzn_B\left(z\right)\left(\frac{1}{z-i\omega+\omega_{k-q}}\frac{1}{z-\omega_q}\right)\right]\notag\\
=& 0
\end{align}
(\ref{final1}) can be rewritten as follows
\begin{align}
\label{final2}
I=R_1+R_2=0
\end{align}
where $R_1=-\sum_1^a{\left(k,w\right)}$, the residue on poles of $n_B$, and $R_2$ is the residue on poles of $G_0$'s. After analytic continuation, the expression of $ \sum_1^a{\left(k,w\right)}$ is
\begin{widetext}
\begin{align}
\label{final3}
{\sum}_1^a\left(k,\omega\right)=\sum_q \frac{\left|\Gamma_1\left(q,k\right)\right|^2}{2}\left\{n_B\left(\omega_k\right)\frac{1}{\omega_q+\omega_{k-q}-\omega-i\delta}+\left[1-n_B\left(\omega_{k-q}\right)\right]\frac{1}{\omega+i\delta-\omega_{k-q}-\omega_q}\right\}
\end{align}
\end{widetext}
At the zero temperature,$n_B=0$, its expression is as follows
\begin{align}
\label{haha7}
{\sum}_1^a\left(k,\omega\right)=\frac{1}{2}\sum_q\frac{\left|\Gamma_1\left(q,k\right)\right|^2}{\omega-\epsilon_q-\epsilon_{k-q}+i\delta}
\end{align}
The other terms due to Wick's theorem can be obtained similarly. The final expression of ${\sum}^a\left(k,\omega\right)$ is
\begin{align}
\label{haha8}
{\sum}^a\left(k,\omega\right)=2\sum_q\frac{\left|\Gamma_1\left(q,k\right)\right|^2}{\omega-\epsilon_q-\epsilon_{k-q}+i\delta}
\end{align}

\bibliography{SWT}

\end{document}